# Aharonov-Bohm oscillations in bilayer graphene edge state Fabry-Pérot interferometers

Hailong Fu[1,2], Ke Huang[1], Kenji Watanabe[3], Takashi Taniguchi[4], Morteza Kayyalha[5], Jun Zhu[1,6*]

**Affiliations**

[1]Department of Physics, The Pennsylvania State University, University Park, Pennsylvania 16802, USA.

[2]Zhejiang Province Key Laboratory of Quantum Technology and Device, Department of Physics, Zhejiang University, Hangzhou, China.

[3]Research Center for Functional Materials, National Institute for Materials Science, 1-1 Namiki, Tsukuba 305-0044, Japan.

[4]International Center for Materials Nanoarchitectonics, National Institute for Materials Science, 1-1 Namiki, Tsukuba 305-0044, Japan.

[5]Department of Electrical Engineering, The Pennsylvania State University, University Park, Pennsylvania 16802, USA.

[6]Center for 2-Dimensional and Layered Materials, The Pennsylvania State University, University Park, Pennsylvania 16802, USA.

*Correspondence to: jzhu@phys.psu.edu (J. Zhu)

**The charge and exchange statistics of an elementary excitation manifest in quantum coherent oscillations that can be explored in interferometry measurements. Quantum Hall interferometers are primary tools to uncover unconventional quantum statistics associated with fractional and non-Abelian anyons of a two-dimensional system, the latter being the foundation of topological quantum computing [1-4]. Graphene interferometers offer new avenues to explore the physics of exotic excitations due to their relatively small charging energies and sharp confinement potentials [5,6]. Bilayer graphene possesses a true band gap to facilitate the formation of quantum confinement [7] and exhibits the most robust even-denominator fractional quantum Hall states that may host non-Abelian anyons [8-11]. Here we present the design and fabrication of a split-gated bilayer graphene Fabry-Pérot interferometer and experimental evidence of Aharonov-Bohm interference at multiple integer quantum Hall states. The versatility of the device allows us to study a wide range of scenarios, determine the velocities of edge states, and assess dephasing mechanisms of the interferometer. These results pave the way to the quest of non-Abelian statistics in this promising device platform.**

The fractional quantum Hall (FQH) effect of a two-dimensional system supports a plethora of many-body phenomena accompanied by exotic low-energy collective excitations [12]. FQH states with an odd-denominator host fractionally charged particles that obey fractional (anyonic) statistics, instead of the conventional fermionic or bosonic statistics [2-4]. Even-denominator FQH states are predicted to host Majorana zero modes, the non-Abelian exchange statistic of which is the foundation of fault-tolerant quantum computing [1,4,13]. The charge and statistics of a quasiparticle have distinct signatures in an interferometry setup [14-17], which have motivated numerous experiments in high-quality GaAs 2D systems [2,18-25]. Using a growth sequence designed to reduce the charging energy of an interferometer, recent experiments in GaAs have obtained evidence of fractional charge and fractional statistics [2,25]. Dual-gated graphene quantum Hall interferometers incorporating thin hexagonal boron nitride (h-BN) dielectric layers have two natural advantages; The first is a small charging energy due to the nearby screening gates. The second is a relatively sharp edge confinement potential that reduces the effect of edge reconstruction and neutral modes [26-33]. Indeed, Aharonov-Bohm oscillations at integer quantum Hall (IQH) states were observed recently in two monolayer graphene-based interferometers [5,6]. However, because monolayer graphene lacks a native band gap, the construction of quantum confinement in an interferometer utilizes the many-body insulating state at filling factor $\nu = 0$ [34,35], which harbors low-energy excitations that may cause decoherence [36]. Bernal-stacked bilayer graphene (BLG), on the other hand, has an electric-field-induced band gap [7,37] and exhibits the strongest even-denominator FQH states with gap energies of a few Kelvins [8-11], as well as the possibility of several other non-Abelian states [11,38]. The development of BLG-based quantum Hall interferometry will be instrumental to the explorations of these fascinating many-body phenomena.

In this work, we report on the design and fabrication of a bilayer graphene Fabry-Pérot interferometer (FPI). Quantum point contacts (QPCs) of our interferometers are formed by gapped BLG through dual gating. Using a dual-split design, we control the carrier density inside and outside the interferometer with a single top gate to ensure density uniformity. We observe Aharonov-Bohm (AB) oscillations as a function of the magnetic field at filling factors $\nu = 2, 3, 4, 8$ and control the interferometer area using a side gate. We determine the velocity of edge states at different filling factors and examine dephasing mechanisms using the temperature and dc-bias dependence of the oscillations. Our work opens the door to using quantum Hall interferometry and related mesoscopic structures to probe correlated phenomena and their elementary excitations in this high-quality, device friendly 2D material.

**Design and characteristics of the Fabry-Pérot interferometer**

Our FPI devices are built on high-quality h-BN/BLG/h-BN stacks made by dry transfer and encapsulated by top and bottom graphite gates. Figs. 1(a) and (b) show an optical micrograph of device 804 and a 3D schematic of the interferometer structure. The device consists of 8 gates. The bottom gate (BG) covers the entire interferometer area except for regions near the contacts, which are doped by the Si backgate to ensure good contacts [39]. The top gating structure, which consists of four split gates (1U, 1D, 2U, 2D), a center gate, and a side gate, is constructed from one contiguous graphite sheet through reactive ion etching, with a trench width of approximately 35 nm (Fig. S1(c) of the Supplementary Information (SI)). Details of fabrication are given in Section 1 of the SI. The FPI consists of two QPCs, each with an opening of $d$ = 165 nm. We define the QPC by opening a band gap of 20-30 meV in the four dual-gated regions (1U, 1D, 2U, and 2D) and place the Fermi level at mid gap. This design is different from monolayer graphene interferometers reported in Refs. [5] and [6], where a correlated insulator at $\nu = 0$ is used to form the confinement. The $\nu = 0$ state in mono and bilayer graphene is known to harbor gapless magnons, thus its immediate vicinity to edge states raises potential concerns for dephasing [36,40,41]. Using gapped BLG to form the QPCs eliminates this possibility. The dual-split design shown in Fig. 1 enables us to use a contiguous center gate (CG) to control the filling factor $\nu$ in the entire device and maintain screening in the opening of the QPC. Our finite element simulations shown in Fig. 1(c) and measurements shown in Fig. 2(a) verify that a uniform carrier density distribution inside and outside the FPI is achieved while the edge states are confined to the trench region and the interferometer loop is as shown in Fig. 1(g). We control the edge state backscattering amplitude by adjusting the size of the QPC opening $d$ in device design. The side gate (SG) adjusts the area of the interferometer $A$ in two ways. When the filling factor underneath the gate $\nu_{sg}$ is smaller than the bulk $\nu$, $A$ increases gradually with the increase of $V_{sg}$. When $\nu_{sg} \geq \nu$, the entire SG area is added to the interferometer loop, resulting a step increase of $A$. We estimate the lithographically defined FPI area in device 804 to be $A$ = 3.6 μm$^2$/3.1 μm$^2$ with/without the SG area. We follow established practices [41-43] to characterize the gates. The top/bottom h-BN thickness is 23 nm/18 nm, which gives rise to a gating efficiency of 7.25 × $10^{11}$ cm$^{-1}$V$^{-1}$/9.26 × $10^{11}$cm$^{-1}$V$^{-1}$respectively. The Coulomb charging energy $E_c$ is estimated to be ~ 10 μeV for $A$ = 3 μm$^2$. The small $E_c$, which is in part due to the two nearby screening graphite gates, is a natural advantage of graphene interferometers [5,6]. It facilitates the observation of Aharonov-Bohm oscillations by suppressing the effect of Coulomb charging that has plagued many prior studies in GaAs [19-23]. $E_c$ of similar magnitude is only achieved in GaAs devices recently using screening wells embedded in the Molecular Beam Epitaxy growth [2,25]; it played a critical role in the observations of the fractional statistics at $\nu = 1/3$.

Figures 1(d)-(g) show schematically the four gating configurations we used to characterize the bulk BLG and the edge state backscattering rate at QPCs 1 and 2. In Fig. 1(d), all six top gates are swept together to measure the magneto-transport through the entire BLG. Fig. 1(h) shows an example of the longitudinal and Hall resistances $R_{xx}$ and $R_{xy}$ at $B$ = 9 T, where the IQH states $\nu = 2, 3, 4$ are well developed and the FQH state at $\nu$ = 8/3 is partially developed. The 8/3 state fully develops at higher magnetic field (Fig. S2 of the SI). In Fig. 1(e), QPC 1 is activated using 1U, 1D, SG. Figs. 1(f) and (g) show the activation of QPC 2 and of both QPCs respectively. Figure 1(i) compares the $R_{xx}$, $R_{L1}$, $R_{L2}$, and $R_L$ traces obtained using the four gating configurations respectively at $\nu$ slightly less than 2, i.e. $\nu$ = 2-. While $R_{xx}$ remains zero, $R_{L1,2}$ shows finite resistance due to backscattering at the QPC. We determine the backscattering rate $r_i$ of each QPC through $R_{Li} = \frac{r_i}{1-r_i} R_{xy}$. When both QPCs are activated, the resistance through the entire FPI is expected to be $R_L = \frac{R_{L1}}{1-r_2} + \frac{R_{L2}}{1-r_1} - \frac{R_{L1} \cdot R_{L2}}{R_{xy}}$ or $R_L \approx R_{L1} + R_{L2}$ (Eq. 1) when the backscattering rates are low. As Fig. 1(i) shows, the measured $R_L$ obeys Eq. (1) extremely well. We focus our measurements in this weak backscattering regime, where $r$ varies from ~ 0.3% to ~ 10% at different filling factors (See Fig. S2 for plot similar to Fig. 1(i) for other filling factors). In this regime, we assume that only the innermost edge state corresponding to the highest filling factor is partially backscattered and participates in the interference phenomenon while the outer edge states are fully transmitted. Our observation of a single magnetic field period in the AB oscillations supports this scenario.

**Properties of the Aharonov-Bohm interference at integer filling factors**

We proceed to investigate AB oscillations at IQH states $\nu$ = 2-, 3-, 4- and 8-. Figure 2(a) plots an overview of $R_{xx}$ (using Fig. 1(d)) and $R_L$ (using Fig. 1(g)) respectively at $T$ = 21 mK and $B$ = 9 T, with regions of interest circled in the plot. Sweeping the magnetic field $B$ slowly we find $R_L$ oscillations periodic in $B$, an example of which is shown in Fig. 2(b) at $\nu$ = 2-. Th fast Fourier transform (FFT) of the trace yields a *B*-field period of $\Delta B$ = 1.57 mT. An important criterion of AB oscillations examines the slope of the stripes in the so-called "pajama plot", which is a two-dimensional color map of $R_L(B, V_{sg})$ as a function of *B* and the side gate voltage $V_{sg}$. In an FPI, the side gate $V_{sg}$ tunes the AB phase by changing the area of the interferometer, $\delta\varphi = 2\pi B \Delta A / \phi_0$, where $\phi_0 = h/e$ is the flux quantum and *A* is the effective area of the interferometer. A more negative side gate $V_{sg}$ decreases the area of our electron interferometer, i.e. $dA/dV_{sg}$ > 0. Thus a constant AB phase of $\delta\varphi = \frac{2\pi}{\phi_0(B\Delta A + A\Delta B)} = 0$ corresponds to $\Delta V_{sg}/\Delta B < 0$, i.e. stripes of a negative slope on the pajama plot. In contrast to AB effect, Coulomb dominated oscillations manifest as positive $\Delta V_{sg}/\Delta B$ stripes on the pajama plot [17]. Figures 2(c) and (d) show two exemplary $R_L(B, V_{sg})$ plots at $\nu$ = 2- and 4- respectively, the negative $\Delta V_{sg}/\Delta B$ slopes of which confirm their AB origin. Similar $R_L(B, V_{sg})$ maps showing the AB oscillations at $\nu$ = 3- and 8- are given in Fig. S3 of the SI. The attainment of the AB regime is facilitated in our devices by the small charging energy $E_c$ due to the large area of the interferometer and close proximity of the top and bottom graphite gates, which are approximately 20 nm away.

The observed AB oscillations appear at multiple integer fillings, in different magnetic fields and over a range of band gap values used to define the QPC's. They are also robust upon thermal cycling. Using measurements and analysis similar to those shown in Figs. 2(b)-(d), we determine the *B*-field period of the oscillations $\Delta B$ and the effective area of the interferometer *A* = $\phi_0/\Delta B$. Table 1 summarizes results obtained at filling factors $\nu$ = 2-, 3-, 4-, and 8- in device 804 while a complete table including results from 804 and 801 at different *D*- and *B*-fields are given in Table 2. At $\nu$ = 2-, we obtain an effective interferometer area of *A* = 2.6 µm$^2$ when the region underneath the SG is set to $\nu_{sg}$ = 0. Setting $\nu_{sg}$ = 3 enlarges the interferometer loop and increases *A* to 3.0 µm$^2$. This situation is illustrated in Figs. 2(e)-(f). The measured interferometer area *A* = 2.6/3.0 µm$^2$ is 87% of the respective areas defined by the midline of the etched trenches in Fig. 1(a). The close match between the measured and lithographically defined dimensions gives us confidence in the trajectory traveled by the edge states and validates the relatively sharp edge confinement potential achieved in our devices (Fig. 1(c)). This will be important in the explorations of the FQH regime, where prior studies have shown that edge state reconstruction occurring in soft confinement potentials produces neutral modes that contribute to the dephasing of the AB interference [26,27,44].

Our results in Table 1 show that from $\nu$ = 2 to $\nu$ = 3, the effective interferometer area *A* decreases by ~ 30% while its change from $\nu$ = 3 to $\nu$ = 4 is negligible. This is consistent with the Landau level structure of BLG [45] and a simple guiding center pictures of the edge states as illustrated in Fig. 2(g)-(h). The gap of $\nu$ = 2 in BLG is much larger than that of $\nu$ = 3 and an area decrease for the innermost edge is expected as $\nu$ increases from 2 to 3. On the other hand, our experimentally obtained *A* at $\nu$ = 8 is 35% larger than that of $\nu$ = 4. This may be due to the expansion of the confinement potential caused by the much higher carrier density at $\nu$ = 8.

Similar to what's observed in monolayer graphene interferometers [5,6], the SG in our devices can modulate the AB oscillations by affecting the area of the interferometer in a wide range of $V_{sg}$. $R_L(V_{sg})$ exhibits oscillations of period $\Delta V_{sg}$ as a function of $\nu_{sg}$. As an example, Fig. S4 of the SI shows the oscillations of $R_L$ ($V_{sg}$) at $\nu$ = 3-, where $\nu_{sg}$ varies from -2 to 4. $\Delta V_{sg}$ generally decreases with increasing $\nu_{sg}$, and is quite small when $\nu_{sg} \geq \nu$. However this trend is not monotonic. The gating of the SG is particularly efficient when $\nu_{sg}$ is an integer, suggesting the screening effect of the BLG underneath the SG plays a role. More discussions on the effect of the SG are given in Section 4 of the SI.

**Dephasing mechanisms of the BLG FPI interferometer**

The AB interference is a phase-coherent phenomenon subject to dephasing mechanisms that may be single electron in origin or arise from interactions with other electrons, impurities and neutral excitations [44,46-49]. Electrons injected into the edge states at energy $\delta\varepsilon = eV_{sd}$ above $E_F$ encloses a slightly different interferometer area $A$, which results in an additional phase $\delta\varphi = 2\pi \frac{2LeV_{sd}}{h\mathrm{v}}$ (Eq. 2) that shifts the max/min of the AB oscillations [14]. Here $V_{sd}$ is a dc bias applied between the source and drain contacts, $L$ the length of the edge state path between the two QPC's, v the edge state group velocity and $h$ the Planck's constant. Eq. (2) appears as a checkerboard-like pattern in a two-dimensional plot of $R_L(V_{sd}, V_{sg})$. Indeed, this behavior is observed in our interferometers at all filling factors, an example of which is shown in Fig. 3(a) for $\nu$ = 2-. A slight tilt of the checkerboard pattern is attributed to the dc bias asymmetry across the interferometer. Eq. (2) does not lead to a decay of the oscillation amplitude of $R_L$ with increasing $V_{sd}$ however our measurements show this behavior clearly, suggesting that electron-electron interaction induced dephasing plays an important role [48]. We extract the oscillation amplitude $\mathcal{A}$ of the data in Fig. 3(a) and plot its $V_{sd}$ dependence in Fig. 3(b). Fits to $\mathcal{A} \propto \exp\left(-2\pi\alpha \frac{e|V_{sd}|}{E_{TH}^{dc}}\right)\sqrt{cos^2\left(2\pi \frac{eV_{sd}}{E_{TH}^{dc}}\right) + 4x^2 sin^2\left(2\pi \frac{eV_{sd}}{E_{TH}^{dc}}\right)}$ (Eq. 3) allow us to extract a ballistic Thouless energy $E_{TH}^{dc} = e\Delta V_{sd} \approx 64$ μeV [5]. Here $\Delta V_{sd}$ represents one full period of the amplitude change. $E_{TH}^{dc}$ is related to the edge state velocity v and inter-QPC path length $L$ by $E_{TH}^{dc} = \frac{h\mathrm{v}}{L}$ [14], from which we obtain v $\approx 6.3\times10^4$ m/s using $L$ = 4.0 μm for $\nu$ = 2. Similar measurements and analysis are performed for other filling factors, from which we determine $E_{TH}^{dc}$, the damping factor $\alpha$, and the asymmetry factor $x$. The data are shown in Figs. S5 and S6 of the SI and the parameters are summarized in Table 2. Fig. 3(c) plots the edge state velocity v for $\nu$ = 2, 3, 4 and 8 in device 804. The values of v, and the trend of decreasing v with increasing $\nu$, are similar to what Nakamura et al observed in GaAs devices incorporating the screening wells [25]. Compared to similar measurements in monolayer graphene [5,6], the velocity of the $\nu$ = 2 edge state in BLG is 2-3 times smaller, likely due to the different Landau level structures of the two 2D systems. The $\nu$ = 2 gap in BLG is a broken-symmetry gap while it is a primary Landau level gap in monolayer graphene [12,45].

The increase of temperature leads to the suppression of the AB oscillations, as our data in Fig. 3(d) show. Figure 3(e) plots the $T$-dependent visibility of the oscillations $\mathcal{V}$, defined as $\mathcal{V} = \frac{R_{Lmax} - R_{Lmin}}{R_{Lmax} + R_{Lmin}}$. Our data are consistent with an exponential decay given by $\exp\left(-\frac{T}{T_0}\right)$ [14], the fits to which yield $T_0 \approx$ 26 mK at $\nu = 2$ and $T_0 \approx 18$ mK at $\nu = 3$ and the corresponding thermal Thouless energy $E_{TH}^{T} = 4\pi^2 k_B T_0 = 88$ μeV and 61 μeV respectively. Unlike the edge state velocity v, both $E_{TH}^{T}$ and $E_{TH}^{dc}$ are inversely proportional to the total length of the interferometer such that a direct comparison among devices of different dimensions [5,6,25] is not meaningful. The ratio $E_{\text{TH}}^{\text{T}}/E_{\text{TH}}^{\text{dc}}$, however, is size independent and carries important insights regarding the dephasing sources. $E_{\text{TH}}^{\text{T}}/E_{\text{TH}}^{\text{dc}}$ is 1.3 to 1.4 in our devices; $E_{\text{TH}}^{\text{T}}/E_{\text{TH}}^{\text{dc}}$ close to 1 was also reported in monolayer graphene interferometers [5,6]. It is approximately 2.8 in GaAs devices used by Nakamura et al in Ref. [25]. Understanding the origin of the differences can further elucidate the nature of the dephasing mechanisms active in quantum Hall interferometers implemented in different materials. The demonstration of Aharonov-Bohm interference in a BLG-based quantum Hall interferometer is a significant step towards the studies of fractional and non-Abelian excitations in this versatile device architecture.

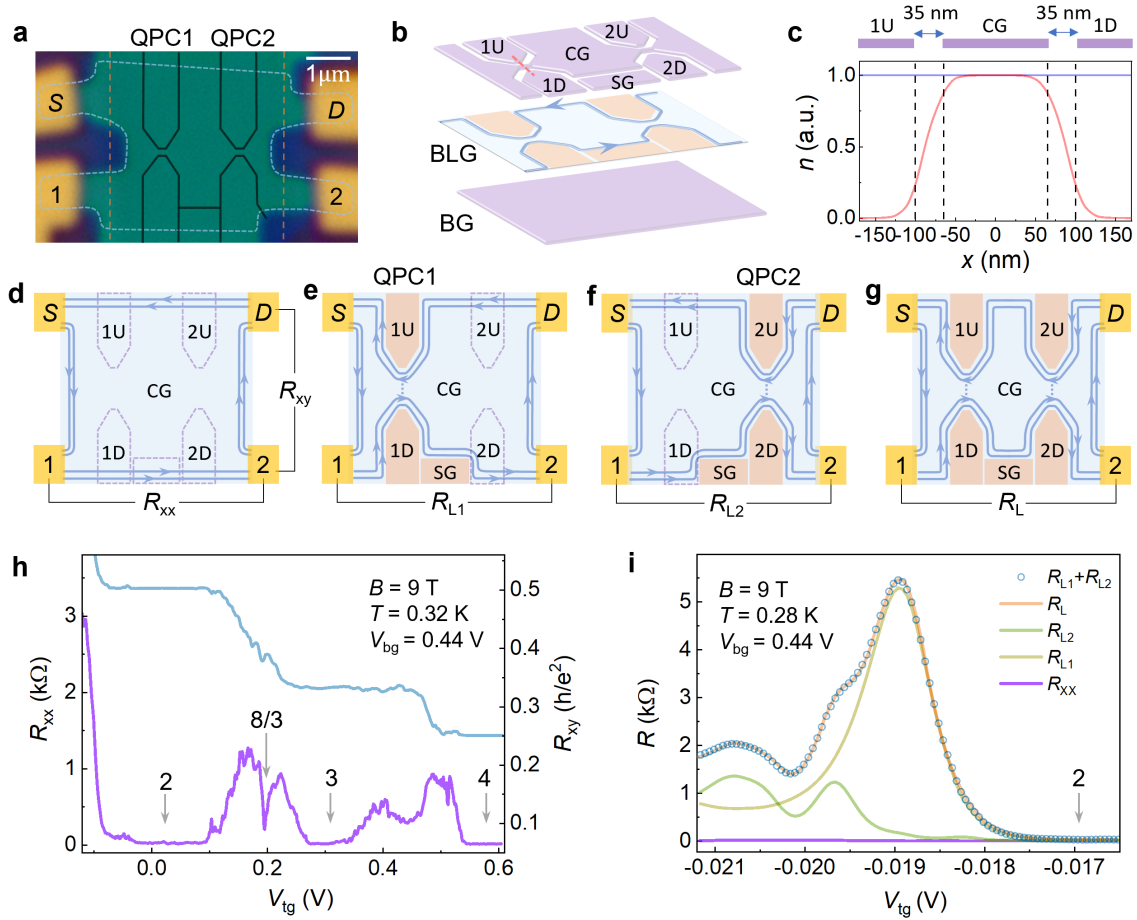

**Figure 1. A bilayer graphene Fabry-Pérot interferometer.** (a) and (b) show an optical micrograph of device 804 and a 3D schematic of the device structure, respectively. The interferometer employs one global graphite bottom gate and six split top graphite gates etched from one contiguous sheet to form the two QPCs, one side gate, and one center gate. The etched trenches are approximately 35 nm in width (black solid lines in (a)). The blue and yellow dashed lines in (a) outline the edges of the BLG sheet and the bottom graphite gate, respectively. A doped silicon backgate not shown here is used to dope the contact areas outside the graphite bottom gate. The QPC opening, defined as the shortest distance between the midlines of the trenches, is 165 nm in device 804 and 135 nm in device 801. See supplementary Fig. S1 for additional images. (c) shows the simulated carrier density profile along the red dashed line in (b). Areas underneath 1U, 1D, 2U, and 2D are depleted. (d)-(g) illustrate the measurement setup and gating configurations used to obtain transport through the bulk BLG, the individual QPCs, and the entire interferometer respectively. Blue solid lines with arrows show the flow of the edge states. $V_{tg}$ denotes voltage applied to the sweeping top gates. In (d), all top gates are swept together. In (e)-(g), the shaded areas are depleted using the respective gates while the remaining top gates are swept together. (h) $R_{xx}$ ($V_{tg}$) and $R_{xy}$ ($V_{tg}$) of the bulk BLG. A contact resistance of 1 KΩ is subtracted from $R_{xy}$. (i) Measured $R_{xx}$, $R_{L1}$, $R_{L2}$, and $R_L$ as a function of $V_{tg}$ and the calculated sum of $R_{L1} + R_{L2}$ near $\nu$ = 2. The excellent agreement between $R_L$ and $R_{L1} + R_{L2}$ indicates weak backscattering of the edge states. $V_{bg}$ = 0.44 V produces a *D*-field of 150 mV/nm in the dual gated regions. From device 804.

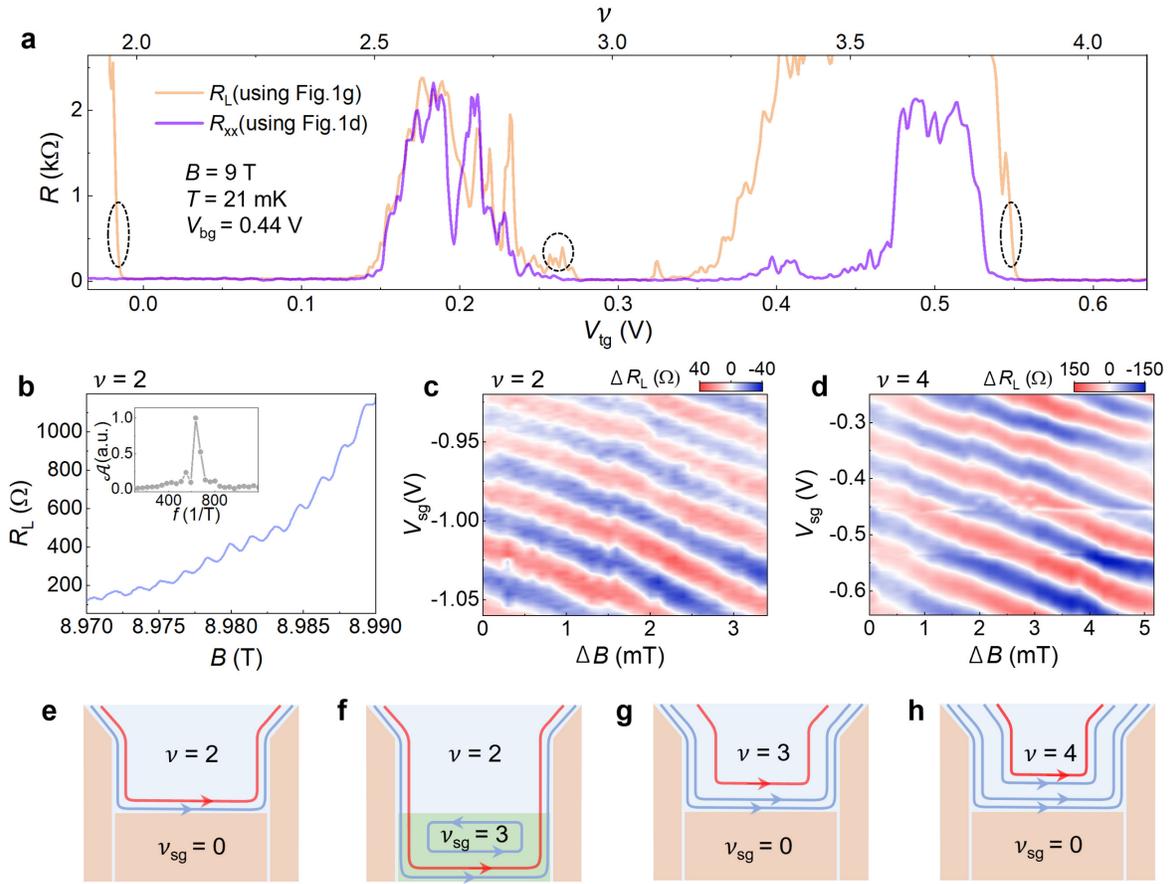

**Figure 2. Aharonov-Bohm interference at IQH states.** (a) Overview of $R_{xx}$ and $R_L$ for $2 < \nu < 4$ at $B = 9$ T and $T = 21$ mK. The estimated electron temperature is around 35 mK. Dashed circles mark locations where Aharonov-Bohm oscillations are studied. They correspond to $\nu$ = 2-, 3-, and 4- respectively. (b) $R_L$-$B$ oscillations at $\nu = 2$. The inset shows the FFT of the trace, from which we obtain $\Delta B = 1.57$ mT and thus an interferometer area of $A = 2.6$ μm$^2$. (c) and (d) show the false color map of $R_L(B, V_{sg})$ at $\nu = 2$ and 4 respectively. A smooth background is subtracted from $R_L$. $\Delta B = B - B_0$, where $B_0 = 8.9786$ T in (c) and $8.9923$ T in (d). The negative slope is evidence supporting Aharonov-Bohm interference. Similar maps at other integer fillings are given in Fig. S3 of the SI. (e)-(h) illustrate the edge state flow with the bulk and the SG area positioned at different filling factors. Only the innermost edge state (red arrowed curve in each figure) experiences weak backscattering and thus interference. From device 804.

| $\nu$ | 2 | 2 ($\nu_{sg}$ = 3) | 3 | 3 ($\nu_{sg}$ = 4) | 4 | 8 |
|---|---|---|---|---|---|---|
| $\Delta B$ (mT) | 1.57 | 1.36 | 2.25 | 1.80 | 2.43 | 1.80 |
| $A$ (μm$^2$) | 2.6 | 3.0 | 1.8 | 2.3 | 1.7 | 2.3 |

**Table 1: Parameters of Aharonov-Bohm oscillations in device 804.**

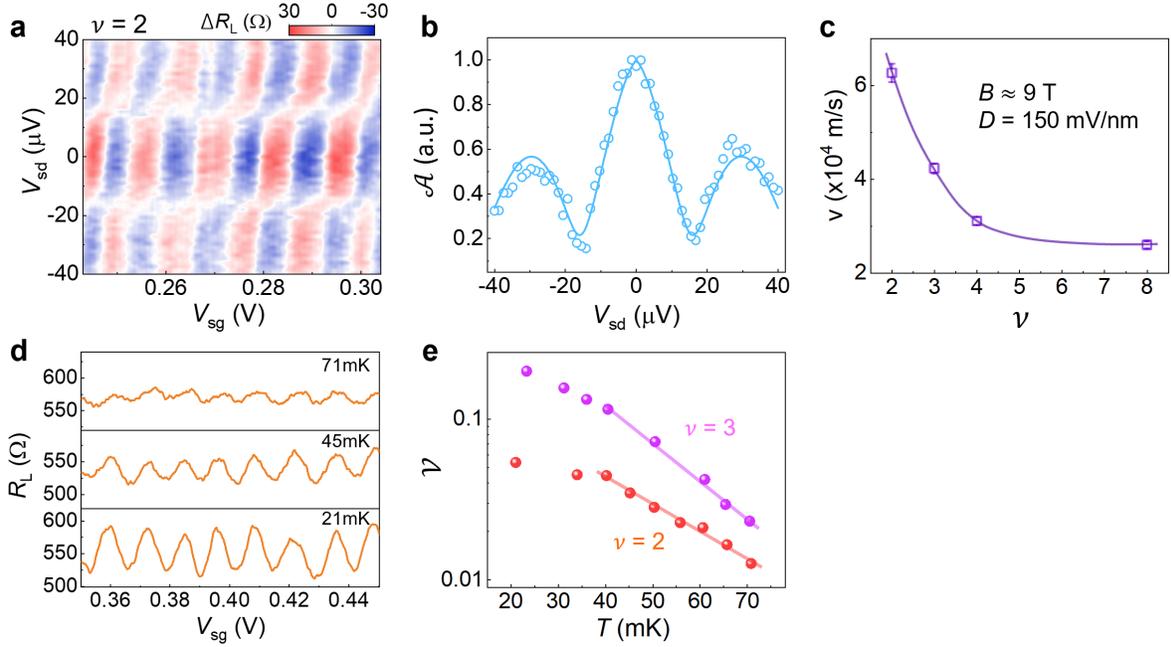

**Figure 3. Edge state velocity and the dephasing of Aharonov-Bohm interference.** (a) shows a false-color map of $R_L(V_{sd}, V_{sg})$ for $\nu$ = 2 at $B$ = 8.9812 T and $T$ = 21 mK. A checkerboard-like pattern is evident, together with a slight tilt of the constant phase stripes. A smooth background is subtracted from $R_L$. (b) plots the oscillations amplitude $\mathcal{A}$ determined from $R_L - V_{sg}$ line scans in (a). The solid line is a fit to Eq. 3, from which we obtain $\Delta V_{sd} \approx 64$ μV, $\alpha = 0.19$ and $x = 0.14$. More discussions, data and fits at other filling factors are given in Section 5 of the SI. (c) plot the edge state velocity at different filling factors in device 804. Here $D$ = 150 mV/nm. The solid line is a guide to the eye. Velocities obtained at other $D$-fields differ slightly. See Table 2. (d) $R_L$-$V_{sg}$ oscillations for $\nu$ = 2 at selected temperatures. (e) plots the visibility $\mathcal{V}$ obtained from traces shown in (d) and similar measurements for $\nu$ = 3. Solid lines are fits to $\mathcal{V} \propto \exp(-\frac{T}{T_0})$, from which we obtain $T_0 \approx 26$ (18) mK for $\nu = 2$ (3). $T_0$ is inversely proportional to the path length $L$ between the QPCs, which is approximately 4.0 μm in device 804.

| device | $\nu$ | $B$ (T) | $D$ (mV/nm) | $A$ ($\mu m^2$) | $L$ ($\mu m$) | $\Delta V_{sd}$ ($\mu V$) | $v$ x10$^4$ (m/s) | damping factor $\alpha$ | asymmetry factor x |
|---|---|---|---|---|---|---|---|---|---|
| 804 | 2 | ~9 | 196 | 2.5 | 4.0 | 58 | 5.6 | 0.07 | 0.14 |
|  | 2 |  | 230 | 2.8 | 4.1 | 60 | 5.9 | 0.05 | 0.04 |
|  | 2 | ~8 | 150 | 2.6 | 4.0 | 64 | 6.3 | 0.19 | 0.14 |
|  | 2 |  |  | 2.7 | 4.0 | 59 | 5.9 | 0.14 | 0.22 |
|  | 3 | ~9 |  | 1.8 | 3.7 | 48 | 4.2 | 0.15 | 0.13 |
|  | 4 |  |  | 1.7 | 3.6 | 36 | 3.1 | 0.22 | 0.16 |
|  | 8 |  |  | 2.3 | 3.9 | 28 | 2.6 | 0.19 | 0.07 |
| 801 | 4 | ~9 | 200 | 1.8 | 4.3 | 42 | 4.4 | 0.32 | 0.16 |

Table 2: Parameters of Aharonov-Bohm oscillations, Thouless energy, edge state velocity, damping factor, and asymmetry factor in bilayer graphene quantum Hall interferometers.

**Acknowledgements**

This work is supported by the Kaufman New Initiative research Grant No. KA2018-98553 of the Pittsburgh Foundation and by the National Science Foundation through the grant NSF-DMR-1904986. H.F. acknowledges the support of the Penn State Eberly Research Fellowship and the ZJU 100 Talents Program (107200*1942222R1, 107200*1942222R3, 107200+1942222R3) of Zhejiang University. M. K. acknowledges the support of the National Science Foundation through the grant OIA- 2040667. K.W. and T.T. acknowledge support from JSPS KAKENHI (Grant Numbers 19H05790, 20H00354 and 21H05233). Work performed at the National High Magnetic Field Laboratory is supported by the NSF through NSF-DMR-1644779 and the State of Florida. We are grateful for helpful discussions with John Chalker, Ady Stern, and Xi Lin. We thank Dr. Elizabeth Green for assisting in measurements at the National High Magnetic Field Laboratory.

**Author contributions**

H. F. and J. Z. designed the experiment. H. F. fabricated the devices and made the measurements. K. H. assisted in device fabrication and measurements. M. K. assisted in measurements. H. F. and J. Z. analyzed the data. K. W. and T. T. synthesized the h-BN crystals. H. F. and J. Z. wrote the manuscript with input from all authors.

**Competing interests**

The authors declare no competing interests.

**Additional information**

Supplementary information is available online.